\documentclass[nohyper,letterpaper,11pt,notoc]{JHEP3}
\usepackage{graphicx}
\usepackage{amssymb}
\usepackage{amsmath}
\usepackage{latexsym}
\usepackage[all]{xy}
\usepackage{slashed}
\preprint{}

\newcommand{\be}{\begin{equation}}
\newcommand{\ee}{\end{equation}}
\newcommand{\bea}{\begin{eqnarray}}
\newcommand{\eea}{\end{eqnarray}}

\newcommand{\TeV}{~\mathrm{TeV}}
\newcommand{\Aa}{A_{\scriptscriptstyle A}}
\newcommand{\Av}{A_{\scriptscriptstyle V}}
\newcommand{\sL}{\scriptscriptstyle L}
\newcommand{\sR}{\scriptscriptstyle R}

\DeclareMathOperator{\SU}{SU}

\DeclareMathOperator{\lie}{Lie}

\DeclareMathOperator{\tr}{Tr}
\DeclareMathOperator{\U}{U}

\title{Anomalies in Fermionic UV Completions of Little Higgs Models}
\author{David Krohn and Itay Yavin~\footnote {dkrohn@princeton.edu, iyavin@princeton.edu} \\ \it{Joseph Henry Laboratories, Princeton University, Princeton NJ 08544}}

\abstract{We consider fermionic UV completions of little Higgs models and their associated $T$-parity-violating anomalous vertices. In particular, we investigate strategies to avoid such parity-violating anomalies. We show that it is unlikely a QCD-like UV completion could be used to implement a model with anomaly-free global symmetry groups. This is because the vacuum state is unlikely to achieve the necessary alignment. However, we will see that certain multi-link moose models, although anomalous, possess a modified form of $T$-parity that leads to a stable particle. Finally, we briefly discuss a discriminant for detecting anomalous decays at colliders.}

\begin{document}
\section{Introduction}
Little Higgs~\cite{ArkaniHamed:2001nc} theories are accorded pride of place among composite models of electroweak~(EW) symmetry breaking.  These models solve the `little' hierarchy problem and are not immediately ruled out by precision EW measurements.  Continuous advances in model building~\cite{Cheng:2004yc,Cheng:2003ju} have given rise to a parity ($T$-parity, analogous to $R$-parity in SUSY) that helps little Higgs theories better satisfy precision EW data by excluding many dangerous tree level interactions. Another welcome consequence of such a parity is the presence of a stable dark matter candidate in the spectrum, the lightest $T$-odd particle (LTP). A recent set of papers~\cite{Hill:2007zv,Hill:2007nz} have shown that quantum anomalies violate $T$-parity by the inclusion of Wess-Zumino-Witten~\cite{Wess:1971yu,Witten:1983tw} terms in the full lagrangian.  While these terms are suppressed, and therefore do not introduce problems with precision data, they render the LTP unstable. One may wonder how generic this instability is in little Higgs models. Is it possible to find UV completions of little Higgs models where the stability of the LTP is not spoiled by anomaly terms? In this short paper we consider several possible means of achieving this. 

In section~\ref{sec:probtpar} we investigate the conditions under which WZW anomalies may be completely removed from a little Higgs theory. The quantized nature of the WZW term leads one to hope that through some discrete choice of model parameters this can be achieved. We begin with models based on anomaly-free global symmetries.  Here we find that in QCD-like UV completions of such models the condensing fermions cannot achieve the desired symmetry breaking pattern due to problems of vacuum alignment.  In addition, we consider moose models and their WZW terms for distinct choices of link direction.  We find that these models will always have anomalous terms, although these may possess a parity. 

The parity of anomalous terms in multi-link moose models is the focus of section~\ref{sec:multilink}.  Here we discuss the parity of WZW terms as a relabeling symmetry of the UV theory.  We consider this parity in the context of a Minimal Moose~\cite{ArkaniHamed:2002qx} like model and show that it can lead to a stable LTP.

Finally, in our appendix we discuss a simple way by which the anomalous vertices can be distinguished at the Large Hadron Collider (LHC).  Here we also summarize some results relevant to computing WZW vertices. 

\section{The problem with T-parity}
\label{sec:probtpar}

Hill and Hill~\cite{Hill:2007zv,Hill:2007nz} recently pointed out that $T$-parity is violated in little Higgs theories by WZW terms~\cite{Wess:1971yu,Witten:1983tw}. They convincingly show that such terms will be present in most little Higgs theories discussed in the literature if one imagines a QCD-like UV completion. In what follows, we explore how general this conclusion is and what sort of structures may give rise to a theory free of WZW terms.

\subsection{Linear UV completions}
The most straightforward way to avoid anomalous vertices in a coset model is to UV complete the theory into a linear sigma model of fundamental scalars.  WZW vertices arise because of anomalies from condensing fermions; remove the fermions and you remove the anomalies.  However, such an approach reintroduces the hierarchy problem composite Higgs theories were created to solve. It is possible to avoid this problem by utilizing a supersymmetric linear sigma model as detailed in Ref. \cite{Csaki:2008se}

\subsection{Anomaly free groups}
Another way to avoid WZW terms is to consider a little Higgs theory with global symmetries that are manifestly anomaly free.  Indeed, models based on the $SO(N)$ and $Sp(N)$ groups have been developed~\cite{Chang:2003un,Cheng:2006ht}, some of which have tree level $T$-parity, and a fermionic UV completion of one such model has been carried out~\cite{Batra:2007iz}. While it is possible to use fermions to implement the UV global symmetry of these theories, whether or not the vacuum will realize the IR coset remains a question of vacuum alignment. {\it In what follows we aim to convince the reader that with a QCD-like theory the vacuum will not align itself into the necessary pattern}.  

\begin{figure}[h]
\begin{center}
\includegraphics[scale=1.0]{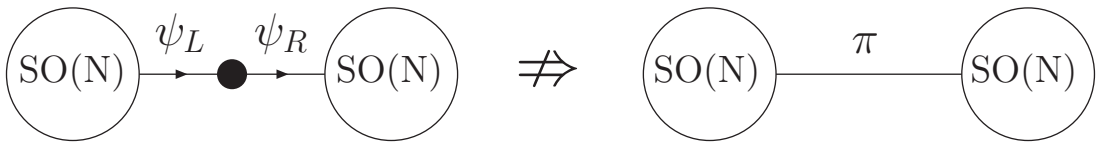}
\end{center}
\caption{A simple-minded attempt to produce a chiral lagrangian with the coset $SO_{\sL}(N)\times SO_{\sR}(N)/SO_{\scriptscriptstyle D}(N)$ from a fermionic QCD-like theory is unlikely to succeed.}
\label{fig:SOxSO}
\end{figure}

For simplicity, consider the coset space $SO_{\sL}(N)\times SO_{\sR}(N)/SO_{\scriptscriptstyle D}(N)$.  The global symmetry of this group is anomaly free; if one could realize this symmetry with fermions then $T$-parity would not be foiled by anomalies.  Here the $L\times R$ structure is needed in order to implement a form of $T$-parity exchanging $L\leftrightarrow R $. A QCD-like UV completion of this model (shown in Fig. \ref{fig:SOxSO}) would consist of quarks transforming as a fundamental and an anti-fundamental, respectively, under some strong gauge group (we take all the fermions to be left-handed Weyl fermions and use the $L/R$-subscripts to designate their position in the moose diagram). As specified, this setup will have a larger global symmetry than we desire: $SU_{\sL}(N)\times SU_{\sR}(N)$. One can try to amend the situation by introducing Majorana masses,
\begin{equation}
\mathcal{L} \supset \psi_{\sL}^{\scriptscriptstyle T} M^{\scriptscriptstyle (L)} \psi_{\sL} + \psi_{\sR}^{\scriptscriptstyle T} M^{\scriptscriptstyle (R)} \psi_{\sR} 
\end{equation}
where $M^{\scriptscriptstyle(L,R)}$ are proportional to the identity in flavor space (we suppress flavor indices to avoid clatter).

We would like the condensate to be $\langle \psi_{\sL} \psi_{\sR}^{\scriptscriptstyle T} \rangle\propto \mathbf{1}$ so as to break the global symmetry to the diagonal subgroup. The low energy theory is then described as usual in terms of the pion fields $U = \exp(2i\pi)$ which span the coset space. Under the global symmetries $U$ transforms like $U\rightarrow L U R^\dagger$, as dictated by the structure of the condensate. We need to choose $M^{\scriptscriptstyle (L,R)} \sim \Lambda_S$ so that $SU(N)$ is strongly broken. Treating the $M^{\scriptscriptstyle (L,R)}$ as a set of spurions transforming as 
\begin{equation}
M^{\scriptscriptstyle (L)}\rightarrow L^\ast M^{\scriptscriptstyle (L)}L^\dagger,\ M^{\scriptscriptstyle (R)}\rightarrow R M^{\scriptscriptstyle (R)}R^T
\end{equation}
we see that the only mass term we can write down for the chiral lagrangian is
\begin{equation}
\mathcal{L}_{mass} = Tr\left( U M^{\scriptscriptstyle (R)}U^T M^{\scriptscriptstyle (L)} \right)
\end{equation}
which indeed gives mass to all the pions associated with the $SU(N)$, but not the $SO(N)$ generators. Raising the mass terms, $M_{ij}^{\scriptscriptstyle (L,R)}\rightarrow \infty$ we decouple all the unwanted goldstones and are left with an $SO_{\scriptscriptstyle L}(N)\times SO_{\scriptscriptstyle R}(N)/SO_{\scriptscriptstyle D}(N)$ coset space. 

However, there is something wrong with this picture. As $M^{\scriptscriptstyle (L,R)}\rightarrow \infty$ all the underlying quarks become heavy and decouple, so how is it that we still have any goldstones left? This is in odds with the persistent mass conjecture~\cite{Preskill:1981sr}: very heavy fermions cannot form a massless goldstone boson. The resolution to this apparent contradiction is that \textit{we are dealing with the wrong goldstones because we have chosen the wrong symmetry breaking pattern}. To see this note that the condensate $\langle \psi_{\sL} \psi_{\sR}^{\scriptscriptstyle T} \rangle\neq0$ is not the only way the vacuum can align itself. The confining strong group must be such that it allows for $\langle \psi_{\sL} \psi_{\sL}^{\scriptscriptstyle T} \rangle\neq0$ and $\langle \psi_{\sR} \psi_{\sR}^{\scriptscriptstyle T} \rangle\neq0$. If this were not the case we would not be able to write the Majorana mass terms to begin with. This new configuration is the correct alignment.  The low energy theory then contains two pions fields $U_{\scriptscriptstyle L,R} = exp(2i \pi_{\scriptscriptstyle L,R})$ each spanning the coset $SU(N)/SO(N)$. It is possible to write a mass term for each independently,
\begin{equation}
 \mathcal{L}_{mass} = Tr\left(U_{\sL} M^{\scriptscriptstyle (L)} + M^{\scriptscriptstyle (L)}U_{\sL}^{\scriptscriptstyle T}\right) \quad +\quad L\rightarrow R
\end{equation}
The paradox is now resolved. As $M^{\scriptscriptstyle (L,R)}\rightarrow \infty$ our pions decouple; none are left in the spectrum. Therefore, adding Majorana masses will not get us the desired symmetry breaking pattern. Indeed, by continuity, this argument seems to imply that the addition of even a small Majorana mass term will misalign the vacuum (although the existence of a phase transition is possible). 

Having put the idea of using fermion masses to achieve the desired symmetry to rest, one could consider trying to enforce an $SO(N)$ global symmetry by adding scalars with a Yukawa coupling to the confining quarks: $y \psi_{\sL}^T h \psi_{\sL}$ \footnote{The scalar can also be charged under the strong group. In that case, the strong group could also be $SU(N)$.}. The Yukawa coupling, $y$ must be very large or else we are only softly breaking the global $SU(N)$ symmetry. Unfortunately, such a setup seems problematic as well. If the scalar's mass is much heavier than $\Lambda_S$, we should integrate it out and generate a 4-fermion operator.  This, however,  will be suppressed and hence constitute only a soft breaking term. Keeping the scalar mass lighter than $\Lambda_S$ will require fine-tuning because of the large Yukawa. This solution will not work without additional ingredients.

Alternatively (or in some sense, equivalently), we can consider 4-fermion operators,
\begin{equation}
\mathcal{L} \supset \frac{y^2}{M^2} \psi_{\sL}^{\scriptscriptstyle T} \psi_{\sL}  \bar{\psi}_{\sL}^{\scriptscriptstyle T} \bar{\psi}_{\sL} \quad +\quad  L\rightarrow R
\end{equation}
Such terms possess a chiral symmetry which forbids fermion masses and the correct alignment of the vacuum seems more plausible. Once again unless we fine-tune $M \sim \Lambda_S$, this term will only lead to a soft breaking of $SU(N)$. However, in analogy with walking technicolor\cite{Holdom:1981rm}, one can imagine a strongly interacting theory which gives rise to large anomalous dimensions for such 4-fermion operators. In that case, the breaking of the global $SU(N)$ can be strong without any fine-tuning. 

Both of the solutions proposed in the last two paragraphs (a finely tuned scalar or a strong theory with operators of large anomalous dimension) seem difficult to implement in standard QCD-like theories, and to the best of the authors' knowledge no realistic examples of these mechanisms are known. However, if one considers {\it supersymmetric} QCD-like theories, then the situation is considerably more hopeful. Indeed, one can then naturally stabilize the scalar or, alternatively, have operators with large anomalous dimensions (such as the gauge duals of fermions in the bulk of AdS). It may be interesting to construct an explicit example of such a theory.

Although we have not proven a no-go theorem, we hope we have convinced the reader of the following: it seems unlikely that a \textit{natural, non-supersymmetric} strongly coupled theory can give rise to a chiral lagrangian with a coset space of $SO_{\sL}(N)\times SO_{\sR}(N)/SO_{\scriptscriptstyle D}(N)$. Similar considerations apply to any other global group with only real representations, e.g. $Sp(N)$ groups. A counterexample to this conclusion would constitute a very welcome addition to the model builder toolkit. 

\subsection{Anomalies in Moose Models}

In light of the preceding discussion, to consider fermionic UV completions it seems natural to work with $\SU(N)$ moose models. If we ignore problems with the Higgs quartic coupling~\cite{Lane:2002pe}, such moose theories are easy to UV complete\footnote{The problem of generating a large quartic coupling in such theories is by no means simple. In \cite{Lane:2002pe}, the author cogently argues  that one will not generate a sufficiently large quartic in theories based on deconstruction. The solution offered in \cite{Gregoire:2002aj} relies on having large number of sites and the authors find that the EW scale is parametrically $v^2\sim f^2/N^2$, where $f\sim 1\TeV$ is the ``pion'' decay constant and $N^2$ is the total number of sites (two extra dimensions). However, this scaling is essentially the same (albeit in one additional extra dimension) as the one worked out in the original little Higgs paper \cite{ArkaniHamed:2001nc}. In such constructions, with $d$ extra dimensions, the EW scale is given by $v^2 \sim f^2/N^d$. Therefore, adding extra dimensions does not help much because parametrically $v^2 \sim f^2/(\text{total \# of sites})$ and in realistic models the number of sites is $\sim \mathcal{O}(1)$. Other ways of generating a large quartic include large Yukawa coupling to matter~\cite{ArkaniHamed:2002qx} or integrating out a heavy scalar with cubic coupling to the higgs \cite{ArkaniHamed:2002qy}.}.  Each link becomes two Weyl fermions condensing at a high scale. If we construct these models with identical strong groups for each link, the only freedom we have in is in selecting the representation of the condensing fermions ($N$ vs. $\bar{N}$), which in turn determines the direction of the link fields. This freedom can be used to cancel gauge anomalies, and one might hope that such arrow adjustments are sufficient to avoid the anomalies violating T-parity.  However, because the WZW terms are sensitive to the global symmetries of a theory they cannot be removed through a choice of link direction.

\begin{figure}[h]
\begin{center}
\includegraphics[scale=1.0]{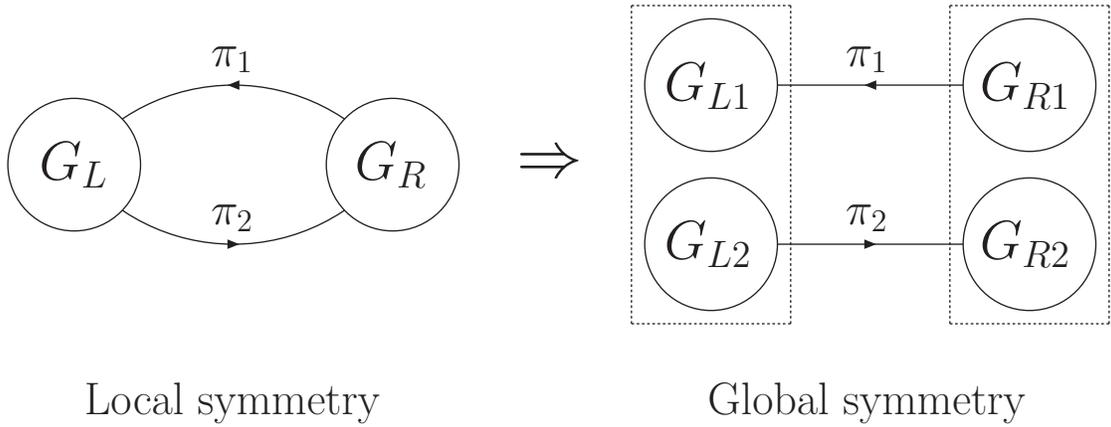}
\end{center}
\caption{The symmetries of the two link moose}
\label{fig:globallocalmoose}
\end{figure}

Let us begin by by considering the action of $T$-parity on a moose model.  In a coset space with the structure $G/H$, and Lie algebras defined as
\begin{equation}
\lie(H)=h,\ \lie(G)=h+k,
\end{equation}
a theory with WZW terms over a symmetric space (one where the commutator of two elements in $k$ lies in $h$) can be split into parity eigenstates as detailed in~\cite{Chu:1996fr}. This parity is defined as the transformation, 
\begin{equation}
\pi\rightarrow-\pi,\ A_h\rightarrow A_h, \ A_k\rightarrow -A_k
\end{equation}
Moreover, in models where  $G=\SU(N)\times\SU(N)$, all WZW terms have negative parity under this transformation~\cite{Chu:1996fr}. We can therefore say that this parity takes
\begin{equation}
\label{eqn:WZWtrans}
\mathcal{L}_{\rm{\scriptscriptstyle WZW}}(\pi, A_h, A_k)\rightarrow-\mathcal{L}_{\rm{\scriptscriptstyle WZW}}(\pi, A_h, A_k)
\end{equation}
Now, for illustration purposes, consider an $\SU(3)$ moose model such as the one considered in ref.~\cite{ArkaniHamed:2002qx} but with only two links for simplicity. This is shown in Fig.~\ref{fig:globallocalmoose}.  We gauge the $\SU(2)\times\U(1)$ subgroup of each $\SU(3)$ where $\SU(2)$ sits in the upper-left hand corner and $\U(1)$ corresponds to the $T_8$ generator. We can schematically write the Lagrangian for this~\footnote{The relative sign between the WZW terms is crucial.  It can be derived by noting that the two $\pi$ fields transform oppositely under the left and right groups, and that the gauged groups here are anomaly free. We thank Hsin-Chia Cheng for pointing out a sign error in an earlier draft of this paper that lead to the wrong conclusion regarding the existence of an LTP.} as
\begin{equation}
\mathcal{L}\sim \mathcal{L}_{\rm{kin}}(\pi_{\scriptscriptstyle 1},A)+\mathcal{L}_{\rm{kin}}(\pi_{\scriptscriptstyle 2},A)+\mathcal{L}_{\rm{\scriptscriptstyle WZW}}(\pi_{\scriptscriptstyle 1},A) + \mathcal{L}_{\rm{\scriptscriptstyle WZW}}(\pi_{\scriptscriptstyle 2},A)
\end{equation}
where $\pi_{\scriptscriptstyle{1,2}}$ are the pions associated with the two links and $A$ are the gauge fields, $A_{\scriptscriptstyle L,R}$ on the left and right sites. The usual definition of $T$-parity takes 
\begin{equation}
\label{eqn:usualT-par}
U_{1/2}\rightarrow\Omega U_{1/2}^\dagger\Omega,\ A_{\scriptscriptstyle L/R}\rightarrow A_{ \scriptscriptstyle  R/L}
\end{equation}
where
\begin{equation}
\Omega=\left(\begin{array}{ccc}1 & 0 & 0 \\0 & 1 & 0 \\0 & 0 & -1\end{array}\right)
\end{equation}
and we have labeled the Goldstones either $d$ (for block-diagonal) or $h$ (because some combination of these will become the Higgs):
\begin{equation}
U=e^{2i\pi/f_\pi},\ \pi\rightarrow\left(\begin{array}{c|c}d & h \\\hline h^\dagger & d\end{array}\right)
\end{equation}
$T$-parity takes $\mathcal{L}_{\rm{kin}}$ to itself. However, as we have just seen, in an $\SU(3)\times\SU(3)/\SU(3)$ model the WZW terms flip their sign under the action of T-parity,
\begin{equation}
 \mathcal{L}_{\rm{\scriptscriptstyle WZW}}(\pi,A_{\sL},A_{\sR}) \stackrel{\text{T-Parity}}{\longrightarrow} - \mathcal{L}_{\rm{\scriptscriptstyle WZW}}(\pi,A_{\sL},A_{\sR})
\end{equation}
An example of this is found in the famous `Cheshire Cat' term with five pion fields that goes to minus itself under $\pi\rightarrow -\pi$. Thus, as pointed out by Hill and Hill~\cite{Hill:2007zv,Hill:2007nz}, such terms violate T-parity as defined in Eq. (\ref{eqn:usualT-par}), independent of the direction of the arrows on the links. This happens because reversing the direction of our links can cancel gauge anomalies, but cannot remove the global anomalies associated with WZW terms.  As can be seen in Fig.~\ref{fig:globallocalmoose}, there are anomalous global symmetries present in moose models.  

Despite this conclusion, the existence of WZW terms does not necessarily forbid a parity of the theory. As we shall see in the next section, when the two links have opposite orientation, a modification of $T$-parity remains intact and ensures a stable particle.

\section{WZW Terms in Multilink Moose Models}
\label{sec:multilink}

There is, however, more to the story of anomalies in multi-link moose models.  Although these models contain WZW terms, {\it when we include multiple links placed in opposite directions we find a parity of the WZW sector!}  A theory with this parity has interesting phenomenological implications, the most striking of which is that the LTP is stable. We will begin by describing the parity of WZW terms defined for symmetric spaces, and then show how an extended version $T$-parity acts to ensure a stable LTP. The example we consider is the two-link model from the previous section, but our arguments can be generalized to physical models with four or more links.

\subsection{$T$-parity}

We can now exploit the previously defined parity in our multilink moose model to define a new $T$-parity under which the full theory is invariant. In this case the direction of the link fields is important. As a simplified example, consider a theory of two links positioned in opposite directions. This theory will have the following kinetic terms,
\begin{equation}
\label{eq:kinsplit}
\begin{split}
\mathcal{L}_{\rm{kin}}(\pi_{1},\pi_{2},A)\quad=&\quad f_\pi^2 \tr\Big|\partial U_{1}-{i}A_{\scriptscriptstyle L}U_{1}+iU_{1}A_{\scriptscriptstyle R}\Big|^2\\
+& \quad f_\pi^2 \tr\Big|\partial U_{2}-{i}A_{\scriptscriptstyle R}U_{2}+iU_{2}A_{\scriptscriptstyle L}\Big|^2
\end{split}
\end{equation}
Defining the vector and axial combinations,
\begin{equation}
A_{\scriptscriptstyle V/A}=\frac{1}{\sqrt{2}}\big(A_{\scriptscriptstyle  L}\pm A_{\scriptscriptstyle R}\big)
\end{equation}
we can write the kinetic term as,
\begin{equation}
\label{eq:kinsplit}
\begin{split}
\mathcal{L}_{\rm{kin}}(\pi_{1},\pi_{2},A)\quad=&\quad f_\pi^2 \tr\Big|\partial U_{1}-\frac{i}{\sqrt{2}}\big[\Av,U_{1}\big] - \frac{i}{\sqrt{2}}\big\{\Aa,U_{1}\big\}\Big|^2 \\
+&\quad f_\pi^2 \tr\Big|\partial U_{2}-\frac{i}{\sqrt{2}}\big[\Av,U_{2}\big] + \frac{i}{\sqrt{2}}\big\{\Aa,U_{2}\big\}\Big|^2
\end{split}
\end{equation}

We identify the antisymmetric pions
\begin{equation}
\pi_{\scriptscriptstyle A}=\frac{1}{\sqrt{2}}(\pi_{\scriptscriptstyle 1}-\pi_{\scriptscriptstyle 2})  = \left(\begin{array}{c|c}d_{\scriptscriptstyle A}& h_A \\\hline h_A^\dagger & d_{\scriptscriptstyle A}\end{array}\right)
\end{equation}
as the light pions whose mass is protected by collective symmetry breaking.  The $d_{\scriptscriptstyle A}$ are eaten by the axial gauge-fields and $h_A$ serves as the SM's higgs doublet. 

Now,  we define $T$-parity as,
\begin{equation}
U_{\scriptscriptstyle 1/2}\rightarrow \Omega U_{\scriptscriptstyle 2/1} \Omega,\ A_{\scriptscriptstyle L/R} \rightarrow A_{\scriptscriptstyle R/L}
\end{equation}
Under this parity, the WZW terms transform into themselves,
\begin{eqnarray}
&\quad& \quad \mathcal{L}_{\rm{\scriptscriptstyle WZW}}(\pi_{1}, \ A_{\scriptscriptstyle L}, \ A_{\scriptscriptstyle R}) + \mathcal{L}_{\rm{\scriptscriptstyle WZW}}(\pi_{2}, \ A_{\scriptscriptstyle R}, \ A_{\scriptscriptstyle L}) \quad \\ \nonumber & \stackrel{T-parity}{\longrightarrow} & \quad \mathcal{L}_{\rm{\scriptscriptstyle WZW}}(\pi_{2}, \ A_{\scriptscriptstyle R}, \ A_{\scriptscriptstyle L}) + \mathcal{L}_{\rm{\scriptscriptstyle WZW}}(\pi_{1}, \ A_{\scriptscriptstyle L}, \ A_{\scriptscriptstyle R})
\end{eqnarray}
so the entire WZW sector is left invariant. This parity guarantees the stability of an LTP. Under this parity, the would be SM Higgs field $h_{\scriptscriptstyle  A}$, as well as the heavy pion $d_{\scriptscriptstyle S}$, are even. The Higgs' partner, $h_{\scriptscriptstyle S}$ is odd and if lighter than the heavy gauge fields, can serve as the LTP. Otherwise, the lightest of the heavy guage-fields is the LTP. 

This modified $T$-parity can be easily generalized to the more realistic four-link moose models that include plaquette operators.  To see this it is instructive to consider the UV perspective of such a theory.  The extended version of $T$-parity we have discussed manifests itself as a relabeling symmetry of the full Lagrangian.  A relabeling of condensing fermions and gauge fields in the UV tells us that a relabeling of Goldstones and gauge bosons must be possible in the IR, guaranteeing the preservation of some form of discrete parity. A forthcoming paper will discuss a more realistic scenario with plaquette operators, along with the issues one encounters when one includes SM fermions.

\section{Conclusions}
\label{sec:Conclusions}
We have investigated both the conditions for, and phenomenology of, WZW terms in little Higgs models with $T$-parity. One way to preserve $T$-parity is through a linear UV completion of the chiral lagrangian into a theory with fundamental scalars (which would likely necessitate supersymmetry). In this paper we explored the possibility of a QCD-like UV completion free of anomalies. We found that unless one resorts to non-standard fermionic UV completions with supersymmetry, or operators with large anomalous dimensions, it is unlikely that anomalous terms can be avoided. Even in moose models with multiple links WZW anomalies cannot be removed.  However, in models with multiple links the WZW terms do possess a slight modification of $T$-parity shared by the entire Lagrangian which permits a stable LTP. 

\textbf{Acknowledgments}: We would like to thank the following people for useful discussions:  Nima Arkani-Hamed, Igor Klebanov, Nathan Seiberg, Jesse Thaler, Herman Verlinde, and Lian-Tao Wang. We would like to extend special thanks to Hsin-Chia Cheng who pointed out a crucial sign error in an earlier draft. Also, we would like to thank the authors of Ref. \cite{Csaki:2008se} for informing us of their own work which takes a complimentary approach to the one investigated herein.

\appendix
\renewcommand{\theequation}{A-\arabic{equation}}
\setcounter{equation}{0}
\section{Distinguishing anomalous vertices through spin measurements}
For models without $T$-parity, the LTP decays quickly to two gauge bosons with a lifetime of order $10^{-15}\rm{s}$~\cite{Barger:2007df}. Such a decay would not leave a displaced vertex. A measurement of the life-time is therefore very difficult without a precise determination of the width which may be smaller than the experimental resolution even for the normal vertex. However, it is possible to distinguish this anomalous vertex from a normal three gauge-boson vertex through a spin measurement. If reconstruction of $\Aa$ is possible, one can form the distribution of the outgoing gauge-bosons, $\Av$, about the axis of polarization. Since $\Aa$ carries unit spin, we expect either $\cos^2\theta$ or $\sin^2\theta$ distributions, depending on the initial polarization of $\Aa$, where $\theta$ is the angle between the outgoing bosons and the axis of polarization. For a normal three gauge-boson vertex we expect a $\sin^2\theta$ ($\cos^2\theta$) distribution if $\Aa$ is transversely (longitudinally) polarized. For the anomalous vertex it is precisely the opposite behavior as is easily seen by angular momentum conservation in the rest frame of $\Aa$. This is summarized in Table \ref{tab:AVVdist}. The initial polarization of $\Aa$ can be established to be longitudinal if it is the product of a heavy fermion decay for example. Such measurements should help unravel the anomalous nature of the vertex.

\begin{table}[h!]
\begin{center}
\begin{tabular}{c|cc}
\hline \\[-2.8ex]
&\multicolumn{2}{c}{\Large{\mbox{$\frac{d\Gamma_{\scriptscriptstyle A\rightarrow VV}}{d\cos\theta}$}}} \\
\raisebox{2ex}{$\Aa$ polarization} & Regular & Anomalous\\ [.7ex]
\hline
\hline
Transverse	& \large{\mbox{$\sin^2\theta$}} & \large{\mbox{$\cos^2\theta$}} \\
\\
Longitudinal	& \large{\mbox{$\cos^2\theta$}} &  \large{\mbox{$\sin^2\theta$}} \\
\hline
\end{tabular}
\end{center}
\caption{The angular distribution of the two outgoing SM gauge-boson about the polarization axis in the rest frame of the heavy vector-boson is a good discriminant between the regular three gauge-boson vertex and the anomalous one.}
\label{tab:AVVdist}
\end{table}

\section{WZW terms for general $G/H$ chiral Lagrangians}
\label{app:WZW}
In this section we will review the motivation for the WZW term (following~\cite{Weinberg:1996kr}) and give a prescription for computing it in the general $G/H$ case as detailed in~\cite{Chu:1996fr}.  We will make an effort to keep it as explicit as possible by including normalization factors, factors of $i$, and dispensing with the language of differential forms.

WZW terms can be thought about as coming from the requirement of anomaly matching as given by 't Hooft.  Begin by considering a global symmetry $G$ that is linearly realized by colored fermions far in the UV.  Here, one could imagine trying to weakly gauge $G$ if there was an additional uncolored spectator sector keeping $G$ anomaly free.  As one goes from the UV into the IR and the colored group becomes confining, the condensate breaks $G$ down to $H$. The theory's degrees of freedom change and $G/H$ is realized non-linearly by Goldstones. Yet, as the fundamental theory preserves gauge symmetry, so should the low energy effective theory.  The Goldstones must reproduce the anomaly of the confined quarks to cancel the contribution to the anomaly from the spectator sector.  {\it Therefore, WZW terms are added to a Lagrangian in order to reproduce the quarks' anomalies in the Goldstone sector}.

Before we write down the anomaly terms we should note a distinction that arises when dealing with anomalies.  To calculate an anomaly one must make a choice of regularization that determines which currents exhibit the anomaly.  Regularizing a theory so that all currents exhibit an anomaly in the same way leads to the so called {\it symmetric} anomaly.  Regularizing so that the unbroken subgroup $H$ is anomaly free leads to the {\it covariant} anomaly.  In this paper we are interested in the case where $H$ is an anomaly-free vector subgroup of $G$. To keep $H$ anomaly free and unbroken we will be interested in the covariant anomaly. We hope that the following will be useful for anyone attempting to compute the actual anomalous vertices and note that the distinction between the {\it  symmetric} and {\it
covariant} anomalies does lead to a numerical difference in the coefficients of such vertices.
\renewcommand{\theequation}{B-\arabic{equation}}
\setcounter{equation}{0}
\subsection{Computation}
Chu, Ho, and Zumino~\cite{Chu:1996fr} give a prescription for calculating the WZW term in general $G/H$ theories subject to the following conditions: $H$ is anomaly free, reductive, and $\pi_4(G/H)=0$.  Their formalism is more than what is required for the simple case of chiral symmetry breaking models, but it would be useful in studying more general models.  They calculate the WZW term to be 
\begin{equation}
\mathcal{L}_{\rm{WZW}}(\pi)=-i\int_0^1\pi_a G_a(\pi,A_{t})dt\quad+\quad B(0)-B(1)\\
\end{equation}
where the $B$ terms are outside the integral sign and
\begin{align}
A_{t\mu}=e^{-it\pi}A_\mu e^{it\pi}-i(\partial_\mu e^{-it\pi})e^{it\pi}
\end{align}
\begin{align}
\label{eq:Ga}
G_a(\pi,A)=&\frac{i}{24\pi^2}\epsilon^{\mu\nu\rho\sigma}\tr\bigg[T_a\Big(\partial_\mu{A}_\nu\partial_\rho{A}_\sigma -\frac{i}{2}\partial_\mu{A}_\nu{A}_\rho{A}_\sigma \\
\nonumber &+\frac{i}{2}{A}_ \mu \partial_ \nu{A}_\rho{A}_\sigma-\frac{i}{2}{A}_ \mu{A}_ \nu \partial_ \rho{A}_\sigma\Big)\bigg]
\end{align}
\begin{align}
\label{eq:counterterm}
B(t)=&\frac{1}{48\pi^2}\epsilon^{\mu\nu\rho\sigma}\tr\bigg[\frac{1}{2}\Big(A_{t\mu}^hA_{t\nu}-A_{t\mu}A_{t\nu}^h\Big)\Big(F_{t\mu\nu}+F^h_{t\mu\nu}\Big)\\
\nonumber &+iA_{t\mu}A_{t\nu}^hA_{t\rho}^hA_{t\sigma}^h+iA_{t\mu}^hA_{t\nu}A_{t\rho}A_{t\sigma}+\frac{i}{2}A_{t\mu}^hA_{t\nu}A_{t\rho}^hA_{t\sigma}\bigg]
\end{align}
Above we have defined $A^h$ to be the restriction of $A$ to $\lie(H)$, $F^h$ to be the field strength tensor formed from $A^h$, and $T_a$ to be a group generator normalized so that $\tr(T_aT_b)=\delta_{ab}$.  In the case of chiral symmetry breaking models it is convenient to write
\begin{equation}
T=\left(\begin{array}{cc}t_1 & 0 \\ 0 & t_2\end{array}\right)
\end{equation}
where $t_1$ and $t_2$ are elements of the Lie algebra transforming left handed Weyl spinors under the two product groups.  

\subsection{Parity in models with chiral symmetry}
In this paper we are interested in the case of chiral symmetry breaking where one Weyl fermion transforms in the $N$ of an $\SU(N)$ and the other transforms in the $\bar{N}$.  The appropriate generators in this case are
\begin{equation}
T^V=\left(\begin{array}{cc}t & 0 \\ 0 & -t^\ast\end{array}\right),\ T^A=\left(\begin{array}{cc}t & 0 \\ 0 & t^\ast\end{array}\right)
\end{equation}
for vector and axial generators, respectively.  Consider the parity on these generators that takes
\begin{equation}
T^V\rightarrow T^V,\ T^A\rightarrow-T^A
\end{equation}
{\it The WZW term of a chiral symmetry breaking model is odd under this parity}.  To show this, take all the terms in $\mathcal{L}_{\rm{WZW}}$ and divide them into parity even and odd parts.  First consider the terms in $G_a$ from eqn.~\ref{eq:Ga}.  Here the generator $T_a$ is in $\lie{K}$, so the remaining generators must contain an odd number of $T^A$ for a given term to be of even parity.  If one makes use of this, combined with the hermiticity of the lie algebra generators, the cyclic properties of the trace, and the antisymmetry of the epsilon symbol one can show that the even terms in $G_a$ vanish for models of chiral symmetry breaking. The proof that Eq.~\ref{eq:counterterm} is odd under this parity proceeds in exactly the same way.

\bibliography{T-par}

\providecommand{\href}[2]{#2}\begingroup\raggedright\begin{thebibliography}{10}

\bibitem{ArkaniHamed:2001nc}
N.~Arkani-Hamed, A.~G. Cohen, and H.~Georgi, {\it {Electroweak symmetry
  breaking from dimensional deconstruction}},  {\em Phys. Lett.} {\bf B513}
  (2001) 232--240, [\href{http://xxx.lanl.gov/abs/hep-ph/0105239}{{\tt
  hep-ph/0105239}}].

\bibitem{Cheng:2004yc}
H.-C. Cheng and I.~Low, {\it {Little hierarchy, little Higgses, and a little
  symmetry}},  {\em JHEP} {\bf 08} (2004) 061,
  [\href{http://xxx.lanl.gov/abs/hep-ph/0405243}{{\tt hep-ph/0405243}}].

\bibitem{Cheng:2003ju}
H.-C. Cheng and I.~Low, {\it {TeV symmetry and the little hierarchy problem}},
  {\em JHEP} {\bf 09} (2003) 051,
  [\href{http://xxx.lanl.gov/abs/hep-ph/0308199}{{\tt hep-ph/0308199}}].

\bibitem{Hill:2007zv}
C.~T. Hill and R.~J. Hill, {\it {$T$-parity violation by anomalies}},  {\em
  Phys. Rev.} {\bf D76} (2007) 115014,
  [\href{http://xxx.lanl.gov/abs/arXiv:0705.0697 [hep-ph]}{{\tt arXiv:0705.0697
  [hep-ph]}}].

\bibitem{Hill:2007nz}
C.~T. Hill and R.~J. Hill, {\it {Topological Physics of Little Higgs Bosons}},
  {\em Phys. Rev.} {\bf D75} (2007) 115009,
  [\href{http://xxx.lanl.gov/abs/hep-ph/0701044}{{\tt hep-ph/0701044}}].

\bibitem{Wess:1971yu}
J.~Wess and B.~Zumino, {\it {Consequences of anomalous Ward identities}},  {\em
  Phys. Lett.} {\bf B37} (1971) 95.

\bibitem{Witten:1983tw}
E.~Witten, {\it {Global Aspects of Current Algebra}},  {\em Nucl. Phys.} {\bf
  B223} (1983) 422--432.

\bibitem{ArkaniHamed:2002qx}
N.~Arkani-Hamed {\em et~al.}, {\it {The minimal moose for a little Higgs}},
  {\em JHEP} {\bf 08} (2002) 021,
  [\href{http://xxx.lanl.gov/abs/hep-ph/0206020}{{\tt hep-ph/0206020}}].

\bibitem{Csaki:2008se}
C.~Csaki, J.~Heinonen, M.~Perelstein, and C.~Spethmann, {\it {A Weakly Coupled
  Ultraviolet Completion of the Littlest Higgs with T-parity}},
  \href{http://xxx.lanl.gov/abs/0804.0622}{{\tt 0804.0622}}.

\bibitem{Chang:2003un}
S.~Chang and J.~G. Wacker, {\it {Little Higgs and custodial SU(2)}},  {\em
  Phys. Rev.} {\bf D69} (2004) 035002,
  [\href{http://xxx.lanl.gov/abs/hep-ph/0303001}{{\tt hep-ph/0303001}}].

\bibitem{Cheng:2006ht}
H.-C. Cheng, J.~Thaler, and L.-T. Wang, {\it {Little M-theory}},  {\em JHEP}
  {\bf 09} (2006) 003, [\href{http://xxx.lanl.gov/abs/hep-ph/0607205}{{\tt
  hep-ph/0607205}}].

\bibitem{Batra:2007iz}
P.~Batra and Z.~Chacko, {\it {Symmetry Breaking Patterns for the Little Higgs
  from Strong Dynamics}},  \href{http://xxx.lanl.gov/abs/arXiv:0710.0333
  [hep-ph]}{{\tt arXiv:0710.0333 [hep-ph]}}.

\bibitem{Preskill:1981sr}
J.~Preskill and S.~Weinberg, {\it {'Decoupling' Constraints on Massless
  Composite Particles}},  {\em Phys. Rev.} {\bf D24} (1981) 1059.

\bibitem{Holdom:1981rm}
B.~Holdom, {\it {Raising the Sideways Scale}},  {\em Phys. Rev.} {\bf D24}
  (1981) 1441.

\bibitem{Lane:2002pe}
K.~Lane, {\it {A case study in dimensional deconstruction}},  {\em Phys. Rev.}
  {\bf D65} (2002) 115001, [\href{http://xxx.lanl.gov/abs/hep-ph/0202093}{{\tt
  hep-ph/0202093}}].

\bibitem{Gregoire:2002aj}
T.~Gregoire and J.~G. Wacker, {\it {Deconstructing six dimensional gauge
  theories with strongly coupled moose meshes}},
  \href{http://xxx.lanl.gov/abs/hep-ph/0207164}{{\tt hep-ph/0207164}}.

\bibitem{ArkaniHamed:2002qy}
N.~Arkani-Hamed, A.~G. Cohen, E.~Katz, and A.~E. Nelson, {\it {The littlest
  Higgs}},  {\em JHEP} {\bf 07} (2002) 034,
  [\href{http://xxx.lanl.gov/abs/hep-ph/0206021}{{\tt hep-ph/0206021}}].

\bibitem{Chu:1996fr}
C.-S. Chu, P.-M. Ho, and B.~Zumino, {\it {Non-Abelian Anomalies and Effective
  Actions for a Homogeneous Space $G/H$}},  {\em Nucl. Phys.} {\bf B475} (1996)
  484--504, [\href{http://xxx.lanl.gov/abs/hep-th/9602093}{{\tt
  hep-th/9602093}}].

\bibitem{Barger:2007df}
V.~Barger, W.-Y. Keung, and Y.~Gao, {\it {T-Anomaly Induced LHC Signals}},
  {\em Phys. Lett.} {\bf B655} (2007) 228--235,
  [\href{http://xxx.lanl.gov/abs/arXiv:0707.3648 [hep-ph]}{{\tt arXiv:0707.3648
  [hep-ph]}}].

\bibitem{Weinberg:1996kr}
S.~Weinberg, {\it {The quantum theory of fields. Vol. 2: Modern applications}},
  . Cambridge, UK: Univ. Pr. (1996) 489 p.

\end{thebibliography}\endgroup
\bibliographystyle{jhep}
\end{document}